\documentstyle[12pt,aaspp4]{article}
\def\hi   {\protect\ion{H}{1}}
\def\oi   {\protect\ion{O}{1}}
\def\hei  {\protect\ion{He}{1}}
\def\hii  {\protect\ion{H}{2}}
\def\oii  {\protect\ion{O}{2}}
\def\sii  {\protect\ion{S}{2}}
\def\nii  {\protect\ion{N}{2}}
\def\oiii {\protect\ion{O}{3}}
\def\hgi  {\protect\ion{Hg}{1}}
\received{}
\accepted{}
\lefthead{Eskridge \& Pogge}
\righthead{\hii\ Regions in NGC 2685}

\begin{document}

\title{\hii\ Region Abundances in the Polar Ring of NGC 2685\footnotemark}
\footnotetext{This paper is partially based on observations obtained with 
the Multiple Mirror Telescope, a joint facility of the Smithsonian Institution
and the University of Arizona.}

\author{Paul B.~Eskridge\footnotemark}
\footnotetext{paul@hera.astr.ua.edu}
\affil{Department of Physics and Astronomy, \break University of Alabama, 
Tuscaloosa, AL 35487}
\and
\author{Richard W.~Pogge\footnotemark}
\footnotetext{pogge@payne.mps.ohio-state.edu}
\affil{Department of Astronomy, \break The Ohio State University, Columbus OH
43210}

\authoremail{paul@hera.astr.ua.edu, pogge@payne.mps.ohio-state.edu}

\begin{abstract}
We have obtained optical spectrophotometry of 11 \hii\ regions in the polar 
ring of NGC 2685 (the Helix galaxy), and have used these data to study the 
physical characteristics of the polar-ring \hii\ regions.  The \hii\ regions 
have normal spectra with no suggestion of unusual density, temperature, 
extinction, or composition.  Semi-empirical calculations yield high oxygen 
abundance estimates (0.8--1.1 $Z_{\odot}$) in all \hii\ regions.  This, along 
with the observed $(B-V)$ color, H$\alpha$ equivalent width, and molecular gas 
properties argue against the current picture in which polar rings form from 
tidally captured dwarf irregular galaxies, and suggests instead that the rings 
are long-lived, self-gravitating structures as predicted by some dynamical 
models.  This would allow the time required for multiple generations of star 
formation, and for the retention of the resulting enriched ejecta for inclusion 
in further generations of star formation.
\end{abstract}


\section{Introduction}

Polar-ring galaxies (PRGs) are unusual objects, typically composed of an 
early-type (E or S0) host galaxy, with a ring of material (stars, gas, and 
dust) orbiting in a plane roughly perpendicular to the major axis of the host 
galaxy.  As such, they are clear examples of galaxy interactions and tidal 
accrection events.  The first major study of the class of PRGs was that of 
\markcite{prg}Whitmore et al.~(1990).  This paper presented data on $\sim$100 
objects, only 6 of which were kinematically confirmed PRGs.  Subsequent work 
has expanded this to a total of eleven objects (\markcite{c92}Combes et 
al.~1992; \markcite{a93}Arnaboldi et al.~1993; \markcite{rc}Reshetnikov \& 
Combes 1994; \markcite{cea}Cox et al.~1995; \markcite{htr}Hagen-Thorn \& 
Reshetnikov 1996).  \markcite{prg}Whitmore et al.~(1990) provides a good 
general overview of the properties of PRGs.  The host galaxies are typically 
gas-poor systems, while the rings are gas-rich (\markcite{rss}Richter, Sackett 
\& Sparke 1994), blue (\markcite{rc}Reshetnikov \& Combes 1994), and dusty 
(\markcite{s83}Schweizer, Whitmore \& Rubin 1983).  PRGs also tend to be much 
more luminous in the far infrared (FIR) than typical early-type galaxies 
(\markcite{rss}Richter et al.~1994).  Based on the \hi\ masses, and optical 
colors of polar rings, the typical model for the ring donor is a tidally 
accreted dwarf irregular (dI) galaxy.  Other possible donors are the outer 
parts of spiral galaxies, and primordial \hi\ gas clouds.

There are three models that have been suggested to explain the apparent
stability of polar rings.  The first of these is due to \markcite{s83}Schweizer
et al.~(1983), who argue that polar rings are {\it not} stable, but can have 
very long decay times.  They conclude that the timescale for the differential 
precession of the ring is on the order of a Hubble time for orbits sufficiently 
near the pole, but very short for orbits inclined by less than 
$\sim$70$^{\circ}$.  The second model, discussed by a number of groups 
(e.g.~\markcite{dc}Dubinski \& Christoudoulou 1994; \markcite{as}Arnaboldi \& 
Sparke 1994; \markcite{ss}Sackett \& Sparke 1990), is that the rings are 
massive, self-gravitating structures that create stable polar orbits for 
themselves by deforming the potential of the host.  The third model is that 
polar rings form in natural `islands of stability':  Polar orbits may be 
inherently stable if the host potential is oblate-triaxial, as suggested by 
\markcite{kr}Katz \& Rix (1992) and \markcite{scd}Steiman-Cameron \& Durisen 
(1988).

A great deal of recent work on PRGs has focussed on their use as probes of the
extent and shape of the dark matter halos of their host galaxies.  These 
studies interpret the kinematics of PRGs in terms of dynamical models, and thus 
assume PRGs to be at least quasi-equilibrium systems.  On the whole, this work 
indicates PRGs to have very large mass-to-light ratios
(e.g.~\markcite{rc}Reshetnikov \& Combes 1994), and substantially flattened 
dark matter halos (e.g.~\markcite{sae}Sackett et al.~1994 and references 
therein).  In a recent paper, \markcite{cna}Combes \& Arnaboldi (1996) argue 
that the kinematic data for NGC 4650A can be fit by a model in which the dark
matter is distributed following the polar-ring isophotes (and {\it not} the
host galaxy).  This is, in a sense, the extreme case of a self-gravitating 
model.

The fundamental questions regarding polar rings are as follows:  How do the
rings form?  Once formed, how long do they last?  One way of addressing these 
questions is to obtain abundance information on polar rings.  All plausible 
candidates for the ring donors are low-abundance objects ($\la 0.3$ Solar).  
For the rings to have high abundances would indicate that either our notions of 
ring formation are wrong, or that polar rings are long-lived, stable, 
self-gravitating structures.  They must not only survive long enough to undergo 
the star-formation episodes required for self-enrichment, they must also be 
massive enough to retain their enriched ejecta.  As polar rings typically have 
numerous \hii\ regions, the easiest method to directly probe their abundances
is nebular spectrophotometry of those \hii\ regions.  We have obtained such 
data for the prototype PRG NGC 2685, and we report our results in this paper.
In \S 2 we provide background on NGC 2685.  In \S 3 we describe our data set 
and reduction techniques.  Our abundance analysis is presented in \S 4.  We 
discuss our results for NGC 2685, and the possible implications for PRGs as a
class in \S 5, and present a summary, along with our plans for continuing this 
project in \S 6.

\section{Background on NGC 2685}

NGC 2685 (Arp 336, also known as ``The Helix galaxy'' and ``The Spindle'') has 
long been known to be an unusual galaxy (\markcite{ha}Sandage 1961).  It is one 
of the two nearest PRGs (the other being NGC 660), and is thus one of the 
easiest to study in detail.  NGC 2685 is a double ring system, with both an 
inner, polar ring, and an outer ring that is co-planar with the host-galaxy 
disk (see the images in \markcite{ha}Sandage 1961).  Unless otherwise stated, 
when we refer to ``the ring'' we will be speaking of the polar ring.  Some 
basic properties of the system are given in Table 1.  The host galaxy and the 
ring are both fairly red.  The system has $M_{H\,I}/L_B \approx 0.3$, similar 
to field dI galaxies (e.g.~\markcite{rh}Roberts \& Haynes 1994).  The \hi\ 
distribution was mapped by \markcite{s80}Shane (1980).  His Westerbork data 
have a final resolution of $\sim$50$''$, and show the \hi\ to be associated 
with the inner and outer rings, with no detected emission from the host galaxy. 
A comparison with the single-dish observations of \markcite{rss}Richter et 
al.~(1994) shows that the Westerbork maps are not missing any significant flux 
from extended emission.  \markcite{h74}Hodge (1974) was the first to point out 
that the ring has a large population of \hii\ regions.  It is unclear if the CO 
$J=1-0$ emission detected by \markcite{tea}Taniguchi et al.~(1990) is coming 
from the ring, or from the host nucleus, however \markcite{wgb}Watson, Guptill 
\& Buchholz (1994) clearly detect CO $J=2-1$ emission from the polar ring, and 
conclude that the ring is rich in molecular gas.  NGC 2685 has no close, 
massive companions.  There are two cataloged dIs within $\sim$30$'$ (UGC 4683 
and MCG+10-13-030).  UGC 4683 has a comparable radial velocity to NGC 2685 
($V_{\odot}=920~{\rm km~s^{-1}}$, \markcite{rc3}de Vaucouleurs et al.~1991, 
hereafter RC3).  There is no published velocity for MCG+10-13-030.  There is a 
loose group of galaxies between $\approx 1^{\circ}\llap.5$ and 
$2^{\circ}\llap.2$ north of NGC 2685, but all these systems have substantially 
higher radial velocities (1288---1518 ${\rm km~s^{-1}}$ -- \markcite{rc3}RC3).  
Thus they do not appear to be physically associated with NGC 2685.

\section{Data Acquisition and Reduction}

We obtained optical spectrophotometry of eleven \hii\ regions in the polar ring 
of NGC 2685 with the MMT Blue Channel Spectrograph on three successive nights 
in November and December 1994.  Table 2 gives the details of the observations.  
The detector was a Loral 3072$\times$1024 CCD.  We used the 500g/mm 
$\lambda_B$=5410\AA\ grating, and a $1'' \times 180''$ slit, resulting in a 
spectral range of 3500\AA\ --- 7100\AA, and a spectral resolution of 
$\sim$5\AA~(FWHM), with a sampling of 1.15\AA\ per pixel.  We binned the data 
by 2 pixels in the spatial direction before read-out, giving a spatial sampling 
of $0''\llap.6$ per binned pixel.  The seeing was $\sim1''$--1$''\llap.5$ 
throughout the run, and the weather was photometric.  Figure 1 shows a 
continuum-subtracted H$\alpha$+[\nii] emission line image of NGC 2685 obtained 
with the Ohio State Imaging Fabry-Perot Spectrometer (\markcite{pea}Pogge et 
al.~1995) in direct imaging mode on the 1.8m Perkins telescope of the Ohio 
Wesleyan and Ohio State Universities in Flagstaff Arizona.  We have marked the 
observed \hii\ regions and the slit positions on this image.

In addition to the ring \hii\ regions, Fig.~1 shows extended H$\alpha$ emission
around the nucleus of NGC 2685.  This emission was discovered by 
\markcite{mhu}Ulrich (1975).  The gas is rotating about the minor axis of the
host, and has a LINER-type spectrum.  The H$\alpha$ morphology of this
material is asymmetric; the emission is much stronger to the NW than to the SE.
This is also obvious from the [\oii] $\lambda$3727\AA~line-intensities shown
by \markcite{mhu}Ulrich (1975).  Her spectra also show that the NW side is the 
approaching side.  This, and our continuum imaging lead us to believe that the
asymmetry is largely due to internal extinction in NGC 2685.

We corrected our data for bias, overscan, and flat-field patterns in the 
standard way, using VISTA (\markcite{vta}Stover 1988).  We prepared the 
slit-illumination frame using IRAF\footnotemark\footnotetext{IRAF is 
distributed by the National Optical Astronomy Observatories, which are operated 
by the Association of Universities for Research in Astronomy, under cooperative 
agreement with the National Science Foundation.}.  We extracted 1-D spectra 
from the images using the spectrum of a bright continuum source as a trace of 
the positional variation along the dispersion axis.  We obtained HeNeAr lamp 
frames before and after each target observation to provide the primary 
wavelength calibrations.  Typically, we had to apply small secondary 
corrections from the night-sky lines on the actual data frames.  We used 20--50 
pixel regions near the ends of the slits to determine the sky subtraction.  We 
used the KPNO tables to correct the spectra for atmospheric extinction.

We obtained repeated observations of the spectrophotometric standards Feige 15, 
Feige 34, and Hiltner 600 throughout each night in order to flux calibrate our 
spectra.  We observed the standards with a $5'' \times 180''$ slit to 
minimize slit-losses.  We emphasize that although the absolute flux calibration 
of our spectra is subject to unknown systematics due to such things as slit 
placement, sky subtraction, slit losses, and tracking errors, the relative line 
fluxes should be accurately determined.  As we obtained spectra at a number of 
position angles, often quite far from the parallactic angle, and over a 
moderate range in air-mass (typically 1.1--1.4), we investigated the 
possibility that our results are compromised by atmospheric dispersion.  From a 
comparison of the relative line fluxes of H$\alpha$ and [\oii] in spectra of 
the \hii\ region R2, obtained over a range of air-masses, we found no evidence 
of the sort of systematic changes in flux ratios that would indicate a 
substantial problem due to atmospheric dispersion.

After flux-calibrating the individual exposures, we measured the centroids of
the bright emission lines, and used these data to correct all exposures of
a given \hii\ region to the same wavelength scale.  The individual exposures 
were then merged, providing total exposure times of between 2 and 7 hours for 
each \hii\ region.  In Figure 2, we present the final spectra for two of our 
targets, demonstrating the full range in quality of our dataset.

\section{Abundance Analysis}

We extracted emission line fluxes and equivalent widths from our spectra using
the VISTA package LINER (\markcite{po}Pogge \& Owen 1993).  As we do not
detect [\oiii] $\lambda$4363\AA~in any of our spectra, we also evaluated 
the continuum rms at $\sim$4350\AA, to determine upper limits for this line.
The recessional velocity of NGC 2685 (see Table 1) puts the \hgi\ 4358\AA\ 
emission line from the Tucson streetlights into the spectral region between 
H$\gamma$ and [\oiii] 4363\AA.  We avoid this region in estimating the upper 
limit on [\oiii] 4363\AA~emission.  We used the following procedure for our 
line-analysis:  

The foreground absorption for NGC 2685 is $A_B = 0.16 ($\markcite{rc3}RC3).  We 
found that our results were generally more self-consistent if we did not 
correct for this explicitly, but instead determined an overall estimate for the 
absorption using the Balmer decrement.  This is a reasonable approach due to 
the small redshift of NGC 2685.  The calculation was complicated by the 
aforementioned \hgi\ 4358\AA~contamination which prevented us from measuring 
H$\gamma$ in our spectra.  However most of our spectra have good detections of 
H$\delta$, giving us at least two Balmer line ratios with which to determine 
the Balmer decrement.  We note that most of our sample have {\it total} 
computed $A_V$'s that are less than the {\it foreground} value of 0.12 mag (for 
$R_V=3.1$) from the RC3.

We used the observed fluxes of the [\oiii] $\lambda$4959\AA, and [\oiii] 
$\lambda$5007\AA\ lines, along with the 3$\sigma$ upper limits to the [\oiii]
$\lambda$4363\AA\ line to determine upper limits to $T_e$ following the 
prescriptions in \markcite{agn2}Osterbrock (1989), and assuming the electron 
density ($N_e$) in the low-density limit.  We then used the flux ratio of the 
[\sii] $\lambda\lambda$6717, 6731\AA\ lines to determine $N_e$ for a range of 
temperatures consistent with the $T_e$ limits.  This was accomplished following 
\markcite{agn2}Osterbrock (1989), but using the updated atomic physics 
calculations of \markcite{s2}Cai \& Pradhan (1993).  We note that the derived 
densities are commensurate with our assumption of the low-density limit in our 
$T_e$ calculations in all cases.

We determined total $A_V$ estimates from the ratios of the Balmer lines, using
Case B recombination, (\markcite{agn2}Osterbrock 1989) for $T_e = 10^4$ K, and 
assuming absorption equivalent widths of 2\AA\ for the Balmer sequence 
(following \markcite{mrs}McCall, Rybski \& Shields 1985).  For a derived
$A_V < 0.05$ no extinction correction was made.  For $A_V \geq 0.05$, we used
the extinction curve of \markcite{rip}Cardelli, Clayton \& Mathis (1989).  In
Table 3 we present the line strengths for all detected lines in all our 
targets.  The line strengths are scaled to the H$\beta$ fluxes.  We give the
measured fluxes before correction for extinction and Balmer absorption 
($f/f_{\beta}$), and after these corrections ($I/I_{\beta}$).  We also give
the measured equivalent widths in \AA, and the statistical signal to noise for
each line.

As we do not detect [\oiii] $\lambda$4363\AA\ in any of our \hii\ region 
spectra, we cannot derive the oxygen abundance directly.  Instead we use the 
semi-empirical $R_{23}$ method (e.g.~\markcite{ep}Edmunds \& Pagel 1984; 
\markcite{de}Dopita \& Evans 1986) to determine the oxygen abundances.  
$R_{23}$ is defined as follows:
$$R_{23} = {{f_{3727} + f_{4959} + f_{5007}} \over f_{H\beta}},\eqno(1)$$
where the $f$'s represent the fully corrected fluxes for the given lines.  As 
the $R_{23}$--oxygen abundance relationship is double-valued, we need to 
determine which branch of the $R_{23}$-abundance curve is appropriate for our 
objects.  Typically, this is done using the ratio of [\nii] $\lambda$6583\AA\ 
to [\oii] $\lambda$3727\AA\ (\markcite{m94}McGaugh 1994, \markcite{r95}Ryder 
1995), where $\log($[\nii]/[\oii]$) > -1$ implies use of the high-abundance 
branch.  This criterion hold for all our sample.  We thus adopt the upper 
abundance branch calibration of $R_{23}$ with $12+\log(O/H)$ of 
\markcite{de}Dopita \& Evans (1986), as presented by \markcite{ok}Oey \& 
Kennicutt (1993).  We present these results, with all other derived quantities 
for our sample, in Table 4.

The derived O/H abundances listed in Table 4 are limited by a systematic 
uncertainty of $\sim0.2$~dex in $12+\log(O/H)$ in the absolute calibration of 
$R_{23}$ (\markcite{ok}Oey \& Kennicutt 1993).  By comparison, the formal 
statistical uncertainties in the derived $12+\log(O/H)$ based on measurment 
errors in $R_{23}$ range from $\sim0.13$~dex for \hii\ regions with the lowest 
signal-to-noise spectra (R8b and R9), to $\lesssim0.05$~dex in the best 
spectra.

\section{Discussion}

The oxygen abundances we derive for the 11 \hii\ regions in our sample are in
the range $8.82 \leq 12+\log(O/H) \leq 8.98$, with an average value of 
$12+\log(O/H) \approx 8.92$ for the polar ring.  Taking $12+\log(O/H)_{\odot} = 
8.93$ (\markcite{ga}Grevesse \& Anders 1989), this means the \hii\ regions in 
our sample have oxygen abundances of 0.8---1.1 $Z_{\odot}$.  We compare this 
with the results of \markcite{zkh}Zaritsky, Kennicutt \& Huchra (1994), who 
obtained similar data for a sample of field spirals.  Our abundance result 
matches the median of their Sc sample, and fits in the range of their Sbc's.  
When combined with other observed properties, as discussed below, this poses a 
substantial problem for understanding the origin and chemical enrichment 
history of the polar ring in NGC 2685.  

As discussed in \S 1, polar rings are typically considered to be debris of 
captured dI galaxies.  Such systems are metal-poor in all studied cases 
(e.g.~\markcite{mbf}Mathews, Boyd \& Fuller 1993).  Other possible sources of 
the ring material (the outer part of a spiral, or primordial \hi\ gas) will 
also be metal-poor.  In fact, there are few, if any, easily accretable sources 
of high-abundance gas in the Universe.  Further, NGC 2685 is {\it not} a 
particularly luminous system (see Table 1).  It has a total $V$-band luminosity 
roughly that of M33 (\markcite{vdb}van den Bergh 1992).  It is therefore very 
difficult to imagine that it could have accreted metal-rich gas from deep in 
the potential well of a more massive spiral disk.  Thus, for the gas to be 
metal-rich, it must have self-enriched, despite the fact that dIs of similar 
overall properties (gas-content and luminosity) have not been able to do so.  
This means that the polar ring surrounding NGC 2685 must be both stable (with a 
lifetime comparable to the Hubble time) {\it and} self-gravitating.  

We have observed only one \hii\ region from the outer ring (R9).  While our
spectrum of this object has the lowest signal to noise in our dataset, it 
appears to have an oxygen abundance similar to those found for the inner ring
\hii\ regions.  We thus see no evidence for significantly different abundances
in the two rings.  We also see no evidence for any azimuthal pattern in our
abundance results.  This argues that the ring gas is likely to be dynamically
well-mixed, and that our results indicate a generally high oxygen abundance in
the ring gas, rather than local enrichment.

Our results thus strongly support the self-gravitating model of PRGs in the
case of NGC 2685.  This is curious, due to the complex appearence of the 
system.  NGC 2685 possesses both an inner (polar) ring, and an outer (planar)
ring.  Further, the appearance of the inner ring is anything but smooth and
relaxed (NGC 2685 has long been known as ``The Helix Galaxy'' and ``The
Spindle'' after all).  These results thus pose the following problem for
dynamical modellers:  Is it possible to construct a stable, self-consistent
model of a self-gravitating ring structure that reproduces the apparent
complexity of NGC 2685?  The work of \markcite{skd}Steiman-Cameron, Kormendy
\& Durisen (1992) indicates that it may be, but certainly the question is far
from resolved.  The most detailed work on dynamical modelling of NGC 2685 is 
that of \markcite{pc}Peletier \& Christodoulou (1993), who modelled the 
two-ring structure, but did not attempt to reproduce the complex appearence of 
the inner ring.  They concluded that a double-ring structure such as that of 
NGC 2685 can be very long lived if the underlying potential is triaxial.  They 
also argue that the ring stars have a mean age of no more than 5--6 Gyr, based 
on the broad-band colors of the ring, and an assumption that the ring is fairly
metal-poor.  Our results argue against this assumption, and thus indicate a 
substantially younger photometric age for the (optical) mean stellar 
population.

We now turn our attention to fitting together the implications of the observed
photometric, chemical, and dynamical properties of the polar ring.  As noted
in \S 2, the \hi\ distribution in NGC 2685 was mapped by \markcite{s80}Shane 
(1980).  Although recent high resolution aperture synthesis observations of the 
\hi\ in NGC 4650A (\markcite{aea}Arnaboldi et al.~1997) reveal that earlier 
VLA (\markcite{vsk}van Gorkom, Schechter \& Kristian 1987) data had seriously 
underestimated the total \hi\ content, we do not believe that this is a serious 
possibility in the case of NGC 2685.  For NGC 4650A, the \hi\ content from 
single-dish measurements (\markcite{rss}Richter et al.~1994) is substantially 
higher than that from the VLA data, and agrees reasonably well with the new 
ATCA data of \markcite{aea}Arnaboldi et al.~(1997).  No such disagreement 
exists between the single-dish and Westerbork results for NGC 2685.  So, 
taking \markcite{s80}Shane's (1980) results for the \hi\ and dynamical mass of 
the polar ring, we find $(M_{HI}/M_{dyn})_{ring} \approx 0.15$.  According to 
\markcite{rh}Roberts \& Haynes (1994), this is typical of an Sm or Im system.  
However, \markcite{pc}Peletier \& Christodoulou (1993) give a $(B-V) \approx 
0.8$ for the polar ring.  This is quite red; \markcite{rh}Roberts \& Haynes 
(1994) find $(B-V) \approx 0.8$ typical of S0/a to Sa galaxies.  Thus, although 
the polar ring has an \hi\ content typical of an Im, it is subtantially redder 
than such galaxies.  

\markcite{pc}Peletier \& Christodoulou (1993) note that their colors may be 
reddened by the dust in the polar ring.  If the true color of the ring is 
assumed to be a typical Im value ($(B-V) \approx 0.4$), then one must assume 
$E(B-V) \approx 0.4$ from internal extinction.  This is quite high, and 
requires a large (but very model dependent -- e.g.~\markcite{dea}Davies et 
al.~1993; \markcite{wtc}Witt, Thronson \& Capuano 1992) dust content, compared 
to that of a typical Im.  Of course abundance is known to correlate with dust 
content in galaxies (e.g.~\markcite{vp}van den Bergh \& Pierce 1990).  Thus our 
abundance results argue for a higher relative dust content than might be found 
in a low abundace dI.  In fact, while the polar ring is obviously dusty (see 
the image in \markcite{ha}Sandage 1961), the FIR properties of NGC 2685 do not 
suggest an unusually high dust content.  \markcite{rss}Richter et al.~(1994) 
find an {\it IRAS} FIR to blue luminosity ratio of $L_{FIR} / L_B \approx 
-0.53$, only slightly higher than the mean values of $-$0.62 and 
$-$0.88$\pm$0.05 derived for samples of S0s by \markcite{tbh}Thronson, Bally \& 
Hacking (1989) and \markcite{enp}Eskridge \& Pogge (1991).  The $L_{FIR} / L_B$ 
for the ring only will, of course, be higher.  \markcite{enp}Eskridge \& Pogge 
(1991) also note that NGC 2685 is \hi\ rich for its FIR flux, compared to a 
sample of S0s.  This may be largely due to the majority of the \hi\ being 
associated with the outer ring.  \markcite{co}Thronson et al.~(1989) derive a
dust mass for NGC 2685 based on its {\it IRAS} flux.  Scaled to our distance 
estimate, this is $M_d \approx 10^5~M_{\odot}$.  If we assume that all the FIR 
originates from the polar ring (unlikely, as the host galaxy has a LINER 
nucleus), this means $M_{dust}/M_{H\,I} \approx 10^{-4}$.  This argues against 
the existence of a vast quantity of dust in the polar ring.  Finally, we note 
that NGC 2685 falls in the midst of the distribution of $L_{FIR} / L_B$ for Im 
galaxies in the sample of \markcite{mi}Melisse \& Israel (1994), but falls at 
the very low end of their distribution for Sc spirals.  In summary, while the 
polar ring is clearly dusty, it is unlikely to have an internal extinction as 
high as $E(B-V) \approx 0.4$

We have measured a total H$\alpha$+[\nii] equivalent width of 12\AA~for the 
polar ring from our H$\alpha$ imaging data.  This is actually a lower limit, as
there is clearly contamination in our continuum from the underlying host 
galaxy.  We use our equivalent width, and the $(B-V)$ color of 
\markcite{pc}Peletier \& Christodoulou to compare the polar ring of NGC 2685
to the spiral galaxy disk models of \markcite{ktc}Kennicutt, Tamblyn \&
Congdon (1994).  We find reasonably good matches for the observed H$\alpha$
equivalent width and $(B-V)$ color with two models.  One has a birthrate 
parameter of $b \approx 0.14$ and exponentially decaying star formation with a 
decay timescale of $\tau \approx 3.2$ Gyr for a \markcite{k83}Kennicutt (1983) 
IMF.  The other has $b \approx 0.09$ and $\tau \approx 2.6$ Gyr for a
\markcite{imf}Salpeter (1955) IMF.  The range of birthrate parameters we
derive from the two good models overlap the \markcite{ktc}Kennicutt et 
al.~(1994) results for Sa and Sab galaxies.  We cannot obtain a reasonable fit 
with the \markcite{s86}Scalo (1986) IMF; the model that comes closest is 
substantially too blue ($(B-V) \approx 0.58$).  If we assume that such a model 
is correct (i.e.~that the measured $(B-V)$ color is strongly effected by 
reddening), then the Scalo IMF model yields a very high birthrate parameter ($b 
\approx 0.7$), and an unconstrained decay time for star formation ($\tau \ga 8$ 
Gyr).  That is, a basically constant rate of star formation over the Hubble 
time.

Finally, we can use our abundance measurement and the ring gas content to 
examine the chemical enrichment history of the polar ring compared with 
typical spiral galaxies.  It is well known that the simple closed-box model
for chemical evolution (e.g.~\markcite{s63}Schmidt 1963) overpredicts the
enrichment for field spiral galaxies (e.g.~\markcite{ssk}Shields, Skillman \& 
Kennicutt 1991).  The closed-box formalism is expressed
$$Z = y~\ln\mu^{-1}, \eqno(2)$$
where $Z$ is the overall or specific abundance in question, $y$ is the yield 
per unit mass of star formation, and $\mu$ is the gas mass fraction.  As noted 
above, the results of \markcite{s80}Shane (1980) give $(M_{HI}/M_{dyn})_{ring} 
\approx 0.15$, or $\ln\mu \approx -1.9$ in typical notation.  Comparing this 
and our oxygen abundance to the field and Virgo cluster spiral samples of 
\markcite{ssk}Shields et al.~(1991), we find that the polar ring is 
substantially more enriched than field galaxies of a similar $\mu$.  The 
abundance is marginally below the simple closed-box model prediction, and 
substantially lower than that observed for Virgo cluster spirals of similar 
$\mu$.  

This comparison has been done ignoring molecular gas.  The CO $J=1-0$ 
observation of \markcite{tea}Taniguchi et al.~(1990) was a single pointing, 
centered on the galaxy nucleus.  It is unclear if the gas they detect is 
associated with the ring or the host galaxy.  \markcite{wgb}Watson et 
al.~(1994) clearly detect CO $J=2-1$ emission from a map made along the long 
axis of the polar ring.  They detect the rotation of the ring in their data; 
the molecular gas is clearly associated with the polar ring.  
\markcite{wgb}Watson et al.~(1994) derive an estimate of $4.5 \pm 0.9 \times 
10^8 M_{\odot}$ for the molecular mass of the polar ring (scaled to our assumed 
distance) assuming the $I_{CO}$--$M_{H_2}$ relationship of \markcite{ys}Young 
\& Scoville (1991), and further assuming the ratio of CO $J=2-1$ to CO $J=1-0$ 
to be $\sim$0.7.  It is clear that the $I_{CO}$--$M_{H_2}$ depends on abundance 
(e.g.~\markcite{ast}Arimoto, Sofue \& Tsujimoto 1996).  However, the ring has 
essentially Solar abundance, and thus this should not be a factor.  It has also 
been shown (e.g.~\markcite{mb}Maloney \& Black 1988) that the excitation 
temperature of the molecular gas (or equivalently, the density of the 
interstellar radiation field) also effects the $I_{CO}$--$M_{H_2}$ ratio.  Thus 
a high excitation temperature of the molecular gas may be inflating the mass 
estimate of \markcite{wgb}Watson et al.~(1994).

We point this out for the following reason:  \markcite{wgb}Watson et al.~(1994) 
derive an estimate of $\log(M_{H_2}/M_{H\,I}) \approx 0.65$.  This is above
\markcite{ys}Young \& Scoville's median value for S0-Sa galaxies, but well
within the range.  \markcite{ys}Young \& Scoville find a few spirals as late as 
Sc with $\log(M_{H_2}/M_{H\,I})$ values as high as $\sim$0.65.  This, despite an
observed $M_{H\,I}/M_{dyn}$ equal to that found for irregulars.  Furthermore, 
if we accept the \markcite{wgb}Watson et al.~(1994) result, this gives 
$(M_{gas}/M_{dyn})_{ring} \approx 0.73$, or $\ln\mu \approx -0.3$.  It seems 
implausible, at best, that a red, Solar abundance ring can have $\sim$75\% of 
its total {\it dynamical} mass in the form of gas.  If we instead simply 
assume that the $M_{H_2}/M_{H\,I}$ ratio is of order unity, we derive $\ln\mu 
\approx -1.2$.  All systems (both in Virgo and the field) with such high $\mu$ 
values in the study of \markcite{ssk}Shields et al.~(1991) have substantially 
sub-Solar oxygen abundances.  Our observed ($\sim$Solar) abundance is roughly 
half a dex higher than these systems, and falls exactly on the prediction from 
the simple closed box model.

We conclude by considering the implications of this work in the context of the 
recent papers of \markcite{gss}Galletta, Sage \& Sparke (1997) and 
\markcite{aea}Arnaboldi et al.~(1997).  As we mentioned above, 
\markcite{aea}Arnaboldi et al.~(1997) present new ATCA \hi\ observations of NGC 
4650A, and conclude that the total \hi\ content of the polar ring is nearly 
twice that indicated by earlier VLA data.  The total \hi\ mass of NGC 4650A is 
$\sim$8$\times$10$^9~M_{\odot}$, comparable to a typical Sb spiral 
(\markcite{rh}Roberts \& Haynes 1994).  \markcite{aea}Arnaboldi et al.~(1997) 
further conclude that the kinematics of the \hi\ reveal evidence for spiral 
structure in the polar ring.  NGC 2685 and NGC 4650A define the extremes of PRG 
morphology.  In NGC 2685, the radius of the polar ring is small compared to the 
optical radius of the host galaxy, while in NGC 4650A, the ring extends out to 
several host optical radii.  Thus while the systems cannot be simply compared, 
they both present enormous difficulties for models that attempt to form polar 
rings out of material that is easily accreted in the Universe today.  

\markcite{gss}Galletta et al.~(1997) present CO $J=1-0$ observations for a 
sample 10 objects from the catalog of \markcite{prg}Whitmore et al.~(1990).  
They supplement this with observations of two other systems (including NGC 
2685) taken from the literature.  They derive H$_2$ mass estimates for their
sample from their CO $J=1-0$ observations, assuming the Galactic calibration of
\markcite{sea}Solomon et al.~(1987).  The \markcite{gss}Galletta et al.(1997)
H$_2$ mass estimates are in the range $10^{8-9}~M_{\odot}$, with 
$M_{H_2}/M_{H\,I}$ typically 0.1 --- 0.7.  While the calibration of H$_2$ mass
with CO line intensity is uncertain at best, the key point is that PRGs tend to
be quite rich in molecular gas compared to dI systems.  Indeed, they approach
$M_{H_2}/M_{H\,I}$ values typical of normal spiral galaxies 
(e.g.~\markcite{s93}Sage 1993), although the overall H$_2$ masses are small
compared to typical spirals.  

The available data, and a number of dynamical models indicate the polar rings
are stable, long-lived structures.  Furthermore, this work, and others that
we have cited above reveal the rings to have physical and chemical properties
unlike those of any plausible gas donor that exists currently.  We speculate 
that polar rings were formed early in the history of galaxy evolution by 
interactions with relative angular momenta that favored capture into inclined
orbits, rather than full mergers.  Thus the properties of such stable, old
polar rings could provide us with information on the nature of the objects that
merged to form normal galaxies in the current Universe.

\section{Summary and Plans for Future Work}

We have presented the results of our spectrophotometric study of \hii\ regions
in the polar ring of NGC 2685.  The oxygen abundances of the 11 \hii\ regions 
in our sample are all in the range 0.8---1.1 $Z_{\odot}$, with no evidence for
any azimuthal abundance pattern.  Thus the polar ring has a gas phase abundance 
typical of Sbc or Sc spirals, despite having an \hi\ content (and mass 
fraction) typical of a dI galaxy.  At first glance this appears to require the 
polar ring to be a long-lived, self gravitating feature, despite its obvious 
complexity.  While this may be the case, a more careful consideration of the 
observed properties of the ring make it clear that the ring cannot be the 
remnant of anything like a current dI galaxy.  It has a $(B-V)$ color as red as 
a typical S0/a or Sa galaxy.  This, combined with its H$\alpha$ equivalent 
width, indicates a starforming history typical of an Sa or Sab spiral.  It has 
molecular gas properties that are more in keeping with an early-type disk 
galaxy than a dI.  Finally, it is one of those rarest of objects that is 
actually well-fit by the simple closed-box chemical enrichment model.

What this seems to require is that the host galaxy accreted a low-luminosity
object that was either already enriched at the time of the interaction, or that
was massive enough (despite its low optical luminosity) to retain the
enrichment products of stellar evolution over the course of the subsequent
history of the system.  There are no obvious field counterparts of either such
object in the Universe today.  

We cannot draw any sweeping conclusions regarding the class of polar-ring 
galaxies based on these results for one object.  Rather, they indicate that a
larger study is warranted.  We have obtained the nessecary spectrophotometry 
for two other kinematically confirmed PRGs, and have the H$\alpha$ imaging 
needed to define spectrophotometric targets for seven of the remaining eight 
currently confirmed PRGs.

\acknowledgments

We take great pleasure in thanking Carol Heller for being a superb MMT 
Operator, and the MMT TAC for awarding us the observing time.  We also thank
Bill Keel and Dimitris Christodoulou for helpful discussion, and the referee 
for several excellent suggestions.  The OSU IFPS instrument project was 
supported by NSF Grants AST-8822009 and AST-9112879.  This research was 
partially supported by the EPSCoR program under grant EHR-9108761.

\newpage

{

\def\tabrule{\noalign{\hrule}}
\def\pz{\phantom{0}}
\def\pb{\phantom{-}}
\def\pd{\phantom{.}}
\ 
 
\centerline{Table 1 -- Basic Properties}
\vskip0.3cm
 
\newbox\tablebox
\setbox\tablebox = \vbox {
 
\halign{\pz\pz#\pz\pz&\pz\pz#\pz\pz\hfil&\hfil\pz\pz#\pz\pz\hfil\cr
\tabrule
\noalign{\vskip0.1cm}
\tabrule
\noalign{\vskip0.1cm}
 
 & & References \cr
\noalign{\vskip0.1cm}
\tabrule
\noalign{\vskip0.2cm}
$\alpha$(1950.0) & $\pb 08^h51^m41{^s}\llap.2$ & 1 \cr
$\delta$(1950.0) & $+58^{\circ}55'30''$ & 1 \cr
$m_B$ & $\pb$12.12 & 1 \cr
$V_{\odot}$ & $\pb$883 ${\rm km~s^{-1}}$ & 1 \cr
D & $\pb$13.4 Mpc & 1,2 \cr
$M_B$ & $-$18.5 & 1,2 \cr
$(B-V)_{host}$ & $\sim$0.9 & 3 \cr
$(B-V)_{ring}$ & $\sim$0.8 & 3 \cr
$S_{H\,I}$ & $\pb34.2 \pm 4.0~{\rm Jy~km~s^{-1}}$ & 4 \cr
$M_{H\,I}$ &  $\sim$1.4$\times 10^9 M_{\odot}$ & 4,2 \cr
$M_{H\,I}/L_B$ & $\sim$0.3 & 4 \cr
$\log(FIR)$ & $-13.50$ & 4 \cr
$L_{FIR}$ & $\sim$1.8$\times 10^9 L_{\odot}$ & 4,2 \cr
$I_{CO}(1 \rightarrow 0)$ & $\pb14.7 \pm 1.2~{\rm K~km~s^{-1}}$ & 5 \cr
$I_{CO}(2 \rightarrow 1)$ & $\pb25.1~{\rm K~km~s^{-1}}$ & 6 \cr
\noalign{\vskip0.2cm}
\tabrule
}
}
\centerline{ \box\tablebox}

\vskip15pt
1) \markcite{rc3}de Vaucouleurs et al.~1991.  2) Distance derived using the
\markcite{yts}Yahil, Tammann \& Sandage (1977) formalism, and 
$H_{\circ}=75~{\rm km~s^{-1}~Mpc^{-1}}$.  3) \markcite{pc}Peletier \& 
Christodoulou (1993).  4) \markcite{rss}Richter et al.~(1994).  5) 
\markcite{tea}Taniguchi et al.~(1990).  6) \markcite{wbg}Watson et al.~(1994).

}
 
{
\tolerance=500
 
\def\tabrule{\noalign{\hrule}}
\def\pz{\phantom{0}}
\def\pb{\phantom{-}}
\def\pd{\phantom{.}}
\ 

\vskip20pt
 
\centerline{Table 2 -- Observing Log}
\vskip0.3cm
 
\newbox\tablebox
\setbox\tablebox = \vbox {
 
\halign{\pz\pz#\pz\pz&\hfil\pz\pz#\pz\pz\hfil&\hfil\pz\pz#\pz\pz\hfil&\pz\pz#\pz
\pz\hfil&\hfil\pz\pz#\pz\pz\hfil\cr
\tabrule
\noalign{\vskip0.1cm}
\tabrule
\noalign{\vskip0.1cm}
 
\ \ UT Date\ \  & Slit PA & Slit Offsets & \hii\ Regions & Total Exposure \cr
 & $^{\circ}$ & $''$E, $\pz\pz''$N & & sec \cr
\noalign{\vskip0.1cm}
\tabrule
\noalign{\vskip0.2cm}
30 Nov.~1994 & 15 & $\pb$34, $-$16 & R1,R4,R5 & 7200 \cr
	     & 55 & $-$30, $\pb$18 & R2,R6,R8a,R8b & 3600 \cr
01 Dec.~1994 & 55 & $-$30, $\pb$18 & R2,R6,R8a,R8b & 3600 \cr
	     & 16 & $-$26, $\pb\pz$0 & R2,R3,R8c,R9 & 9000 \cr
02 Dec.~1994 & 90 & $\pb\pz$0, $\pb$27 & R2,R7 & 9000 \cr
	     & 15 & $\pb$34, $-$16 & R1,R4,R5 & 1800 \cr
\noalign{\vskip0.2cm}
\tabrule
}
}
\centerline{ \box\tablebox}
 
}

\newpage

{
\tolerance=500
 
\def\tabrule{\noalign{\hrule}}
\def\pz{\phantom{0}}
\def\pb{\phantom{-}}
\def\pd{\phantom{.}}
\ 
 
\centerline{Table 3 -- Line Strengths}
\vskip0.3cm
 
\newbox\tablebox
\setbox\tablebox = \vbox {
 
\halign{\pz#\hfil&\hfil\pz#\pz&\hfil\pz#\pz&\hfil\pz
#\pz&\hfil\pz#\pz&\hfil\pz#\pz&\hfil\pz#\pz&\hfil\pz#\pz&\hfil\pz#\pz&\hfil\pz#
\pz&\hfil\pz#\pz&\hfil\pz#\pz\cr
\tabrule
\noalign{\vskip0.1cm}
\tabrule
\noalign{\vskip0.1cm}
 
\ & R1 & R2 & R3 & R4 & R5 & R6 & R7 & R8a & R8b & R8c & R9 \cr
\noalign{\vskip0.1cm}
\tabrule
\noalign{\vskip0.2cm}
[\oii] $\lambda$3727 \cr
${f/f_{\beta}}^1$ & 1.63 & 1.71 & 2.06 & 2.36 & 1.91 & 1.96 & 1.92 & 2.10 & 2.18 & 2.45 & 2.22 \cr
${I/I_{\beta}}^2$ & 1.62 & 2.14 & 1.79 & 2.21 & 1.81 & 1.78 & 1.67 & 1.80 & 2.47 & 1.97 & 2.19 \cr
Eq.~Width$^3$ & 81.30 & 141.2 & 56.66 & 53.50 & 41.04 & 40.06 & 54.47 & 36.38 & 60.78 & 42.15 & 20.51 \cr
$\sigma^4$ & 16.0 & 11.3 & 12.9 & 10.8 & 6.9 & 10.9 & 12.0 & 4.9 & 4.7 & 5.0 & 3.8 \cr
H$\zeta$ $\lambda$3889 \cr
& 0.07 & 0.11 & & & & & & & & & \cr
\ & 0.11 & 0.16 & & & & & & & & & \cr
\ & 3.85 & 10.39 & & & & & & & & & \cr
\ & 2.6 & 1.4 & & & & & & & & & \cr
H$\epsilon$ $\lambda$3969 \cr
& 0.12 & 0.09 & & & & & & & & & \cr
\ & 0.15 & 0.14 & & & & & & & & & \cr
\ & 6.42 & 7.02 & & & & & & & & & \cr
\ & 2.4 & 1.5 & & & & & & & & & \cr
H$\delta$ $\lambda$4102 \cr
& 0.23 & 0.19 & 0.18 & 0.17 & & 0.20 & 0.17 & 0.28 & & & \cr
\ & 0.26 & 0.26 & 0.26 & 0.26 & & 0.28 & 0.26 & 0.37 & & & \cr
\ & 12.46 & 11.37 & 2.97 & 3.41 & & 3.63 & 2.57 & 3.81 & & & \cr
\ & 6.3 & 3.0 & 3.0 & 2.3 & & 2.7 & 4.3 & 1.8 & & & \cr
[\oiii] $\lambda$4959 \cr
& 0.28 & 0.51 & 0.37 & 0.17 & 0.23 & 0.22 & 0.27 & 0.12 & 0.49 & 0.29 & \cr
\ & 0.27 & 0.48 & 0.32 & 0.15 & 0.20 & 0.20 & 0.23 & 0.11 & 0.42 & 0.23 & \cr
\ & 17.93 & 27.68 & 5.13 & 3.19 & 2.90 & 4.49 & 3.27 & 1.41 & 8.36 & 2.15 & \cr
\ & 7.2 & 9.4 & 5.6 & 3.3 & 3.0 & 4.3 & 5.9 & 1.8 & 3.3 & 2.8 & \cr
[\oiii] $\lambda$5007 \cr
& 0.80 & 1.56 & 1.05 & 0.48 & 0.69 & 0.65 & 0.89 & 0.37 & 1.48 & 0.70 & 0.85 \cr
\ & 0.78 & 1.46 & 0.91 & 0.44 & 0.60 & 0.59 & 0.77 & 0.32 & 1.26 & 0.56 & 0.84 \cr
\ & 51.61 & 86.77 & 14.42 & 9.73 & 8.59 & 13.00 & 11.07 & 4.23 & 25.12 & 5.11 & 34.99 \cr
\ & 24.3 & 25.4 & 18.9 & 8.0 & 8.9 & 10.0 & 15.3 & 3.6 & 10.4 & 6.1 & 4.7 \cr
\noalign{\vskip0.2cm}
\tabrule
}
}
\centerline{ \box\tablebox}

\vfill
\eject

\centerline{Table 3 continued}
\vskip0.3cm

\newbox\tablebox
\setbox\tablebox = \vbox {

\halign{\pz#\hfil&\hfil\pz#\pz&\hfil\pz#\pz&\hfil\pz
#\pz&\hfil\pz#\pz&\hfil\pz#\pz&\hfil\pz#\pz&\hfil\pz#\pz&\hfil\pz#\pz&\hfil\pz#
\pz&\hfil\pz#\pz&\hfil\pz#\pz\cr
\tabrule
\noalign{\vskip0.1cm}
\tabrule
\noalign{\vskip0.1cm}

\ & R1 & R2 & R3 & R4 & R5 & R6 & R7 & R8a & R8b & R8c & R9 \cr
\noalign{\vskip0.1cm}
\tabrule
\noalign{\vskip0.2cm}
[\oi] $\lambda$6300 \cr
& 0.05 & & & & & & & & & & \cr
\ & 0.05 & & & & & & & & & & \cr
\ & 4.10 & & & & & & & & & & \cr
\ & 1.4 & & & & & & & & & & \cr
[\nii] $\lambda$6548 \cr
& 0.33 & 0.27 & 0.39 & 0.44 & 0.47 & 0.47 & 0.31 & 0.48 & 0.66 & 0.52 & \cr
\ & 0.32 & 0.21 & 0.34 & 0.39 & 0.38 & 0.43 & 0.27 & 0.41 & 0.45 & 0.41 & \cr
\ & 29.74 & 17.91 & 7.53 & 11.14 & 7.00 & 15.61 & 4.12 & 9.86 & 12.15 & 13.77 & \cr
\ & 12.1 & 6.4 & 7.6 & 6.1 & 5.4 & 7.4 & 6.3 & 2.8 & 5.3 & 4.0 & \cr
H$\alpha$ $\lambda$6563 \cr
& 2.87 & 2.86 & 3.13 & 3.14 & 3.70 & 3.40 & 2.67 & 3.45 & 4.20 & 3.27 & 2.24 \cr
\ & 2.76 & 2.17 & 2.81 & 2.86 & 3.08 & 3.14 & 2.45 & 3.04 & 2.94 & 2.78 & 2.22 \cr
\ & 252.4 & 187.4 & 61.49 & 80.39 & 56.18 & 113.70 & 35.08 & 71.32 & 77.23 & 33.09 & 447.10 \cr
\ & 58.5 & 41.6 & 47.5 & 38.8 & 27.8 & 39.9 & 37.9 & 21.7 & 22.3 & 22.2 & 11.1 \cr
[\nii] $\lambda$6583 \cr
& 0.95 & 0.83 & 1.23 & 1.33 & 1.48 & 1.43 & 0.90 & 1.23 & 1.66 & 1.36 & 1.09 \cr
\ & 0.91 & 0.62 & 1.07 & 1.18 & 1.19 & 1.30 & 0.78 & 1.05 & 1.13 & 1.09 & 1.08 \cr
\ & 80.96 & 54.39 & 24.21 & 34.39 & 22.95 & 48.56 & 11.83 & 25.39 & 30.44 & 13.77 & 217.70 \cr
\ & 26.7 & 18.2 & 23.2 & 20.6 & 18.6 & 22.6 & 23.4 & 11.2 & 11.4 & 11.1 & 5.9 \cr
\hei\ $\lambda$6678 \cr
& & 0.03 & & & & & & & & & \cr
\ & & 0.02 & & & & & & & & & \cr
\ & & 1.90 & & & & & & & & & \cr
\ & & 1.4 & & & & & & & & & \cr
[\sii] $\lambda$6717 \cr
& 0.28 & 0.22 & 0.49 & 0.57 & 0.74 & 0.53 & 0.25 & 0.66 & 0.63 & 0.76 & \cr
\ & 0.27 & 0.16 & 0.43 & 0.51 & 0.59 & 0.48 & 0.22 & 0.56 & 0.42 & 0.61 & \cr
\ & 26.37 & 15.55 & 9.66 & 15.38 & 11.96 & 17.62 & 3.41 & 11.65 & 9.92 & 7.45 & \cr
\ & 7.7 & 6.5 & 8.3 & 7.7 & 8.1 & 8.4 & 6.9 & 5.8 & 5.3 & 5.6 & \cr
\noalign{\vskip0.2cm}
\tabrule
}
}
\centerline{ \box\tablebox}

\vfill
\eject

\centerline{Table 3 continued}
\vskip0.3cm

\newbox\tablebox
\setbox\tablebox = \vbox {

\halign{\pz#\hfil&\hfil\pz#\pz&\hfil\pz#\pz&\hfil\pz
#\pz&\hfil\pz#\pz&\hfil\pz#\pz&\hfil\pz#\pz&\hfil\pz#\pz&\hfil\pz#\pz&\hfil\pz#
\pz&\hfil\pz#\pz&\hfil\pz#\pz\cr
\tabrule
\noalign{\vskip0.1cm}
\tabrule
\noalign{\vskip0.1cm}

\ & R1 & R2 & R3 & R4 & R5 & R6 & R7 & R8a & R8b & R8c & R9 \cr
\noalign{\vskip0.1cm}
\tabrule
\noalign{\vskip0.2cm}
[\sii] $\lambda$6731 \cr
& 0.20 & 0.13 & 0.33 & 0.38 & 0.53 & 0.38 & 0.21 & 0.46 & 0.36 & 0.49 & \cr
\ & 0.19 & 0.09 & 0.29 & 0.34 & 0.42 & 0.34 & 0.18 & 0.39 & 0.24 & 0.39 & \cr
\ & 17.53 & 8.98 & 6.44 & 10.03 & 8.56 & 12.63 & 2.75 & 8.15 & 5.66 & 4.74 & \cr
\ & 5.4 & 2.7 & 5.9 & 5.2 & 6.1 & 7.2 & 4.8 & 4.0 & 3.7 & 4.1 & \cr
\noalign{\vskip0.1cm}
\tabrule
\noalign{\vskip0.2cm}
H$\beta$ $\lambda$4861 \cr
$f_{H\beta}$ & 5.98 & 3.63 & 4.78 & 2.10 & 1.86 & 4.51 & 2.94 & 0.97 & 0.97 & 1.21 & 0.88 \cr
$I_{H\beta}$ & 6.59 & 8.64 & 5.50 & 2.51 & 2.82 & 4.96 & 3.39 & 1.14 & 2.52 & 1.51 & 0.89 \cr
\ & 68.76 & 54.02 & 13.34 & 22.75 & 13.58 & 19.80 & 13.27 & 12.01 & 14.38 & 8.12 & 147.30 \cr
\ & 25.6 & 16.9 & 17.6 & 12.6 & 10.5 & 13.2 & 17.6 & 7.3 & 6.9 & 5.8 & 6.1 \cr
\noalign{\vskip0.2cm}
\tabrule
}
}
\centerline{ \box\tablebox}

1:  The ratio of the uncorrected line flux to that of H$\beta$.  For H$\beta$
the flux is given in units of $10^{-16}~{\rm erg~s^{-1}~cm^{-2}~\AA^{-1}}$.

2:  The ratio of the reddening corrected line flux to that of H$\beta$.  For
H$\beta$ the flux is given in units of $10^{-16}~{\rm 
erg~s^{-1}~cm^{-2}~\AA^{-1}}$.

3:  The measured equivalent width in \AA ngstroms of the line (no absorption 
correction has been applied to the values for the Balmer lines).

4:  The statistical signal to noise of the line.  
}

\newpage

{

\tolerance=500
 
\def\tabrule{\noalign{\hrule}}
\def\pz{\phantom{0}}
\def\pb{\phantom{-}}
\def\pd{\phantom{.}}
\ 
 
\centerline{Table 4 -- Derived Results}
\vskip0.3cm
 
\newbox\tablebox
\setbox\tablebox = \vbox {
 
\halign{\pz\pz#\pz\pz&\pz\pz#\pz&\hfil\pz#\pz\hfil&\hfil\pz#\pz\pz&\pz\pz#\pz
\hfil&\hfil\pz#\pz\hfil&\hfil\pz#\pz\hfil&\hfil\pz#\pz\pz\hfil\cr
\tabrule
\noalign{\vskip0.1cm}
\tabrule
\noalign{\vskip0.1cm}
 
\hii\ Region & Offset$^1$ & $T_e$ & $N_e\pz$ & $A_V$ & $R_{23}$ & 12+log(O/H) 
& Range$^2$ \cr
 & $\pz\pd''$E,$''$N & K & ${\rm cm^{-3}}$ & mag. & & \cr
\noalign{\vskip0.1cm}
\tabrule
\noalign{\vskip0.2cm}
R1 & $\pb$33,$-$22 & $<$11500 & 41 & 0.06 & $\pb$2.67 & $\pb$8.94 & 8.93--8.95 
\cr
R2 & $-$19,$\pb$28 & $<\pd$8700 & $<$10 & 0.77 & $\pb$4.08 & $\pb$8.83 & 
8.81--8.86 \cr
R3 & $-$29,$\pz-$8 & $<$11800 & $<$10 & 0 & $\pb$3.02 & $\pb$8.92 & 8.90--8.93 
\cr
R4 & $\pb$35,$-$12 & $<$15500 & $<$10 & 0.08 & $\pb$2.80 & $\pb$8.94 & 
8.91--8.96 \cr
R5 & $\pb$36,$\pz-$8 & $<$18900 & 41 & 0.26 & $\pb$2.61 & $\pb$8.95 & 
8.92--8.98 \cr
R6 & $-$40,$\pb$13 & $<$13300 & 57 & 0 & $\pb$2.57 & $\pb$8.95 & 8.93--8.97 \cr
R7 & $\pb\pz$5,$\pb$28 & $<$12700 & 232 & 0 & $\pb$2.67 & $\pb$8.94 & 
8.92--8.96 \cr
R8a & $-$30,$\pb$20 & $<$29100 & 25 & 0 & $\pb$2.22 & $\pb$8.98 & 8.93--9.00 \cr
R8b & $-$25,$\pb$23 & $<$15600 & $<$10 & 0.77 & $\pb$4.15 & $\pb$8.82 & 
8.77--8.90 \cr
R8c & $-$23,$\pb$21 & $<$20000 & $<$10 &  0 & $\pb$2.76 & $\pb$8.94 & 
8.88--8.98 \cr
R9 & $\pz-$7,$\pb$62 & & & 0 & $<$3.03 & $>$8.92 & 8.84--8.97 \cr
\noalign{\vskip0.2cm}
\tabrule
}
}
\centerline{ \box\tablebox}

\vskip15pt

1.  With respect to the central position, given in Table 1.

2.  The 1$\sigma$ statistical range in 12+log(O/H) derived from the counting
statistics of the line fluxes.
}

\vskip30pt

\figcaption{NGC 2685 in H$\alpha$ (greyscale) and red continuum (contours).  
The H$\alpha$ emission shows the \hii\ regions in the inner and outer rings, as 
well as the central LINER emission region.  The line-segments show the 
placement of the spectrograph slit.  The labelled circles show the \hii\ 
regions observed.  For clarity, the three observed components of the \hii\ 
complex R8 are not labelled.  They lie between R2 and R6.
}

\figcaption{Fully reduced, flux calibrated spectra of two \hii\ regions in 
NGC 2685, showing the full range in signal to noise of our spectra.  The
spectra shown are from a) R2 with 7 hours total integration, and b) R9 with 2.5
hours total integration.  The spectral features identfied in R2 are labelled in
a).
}

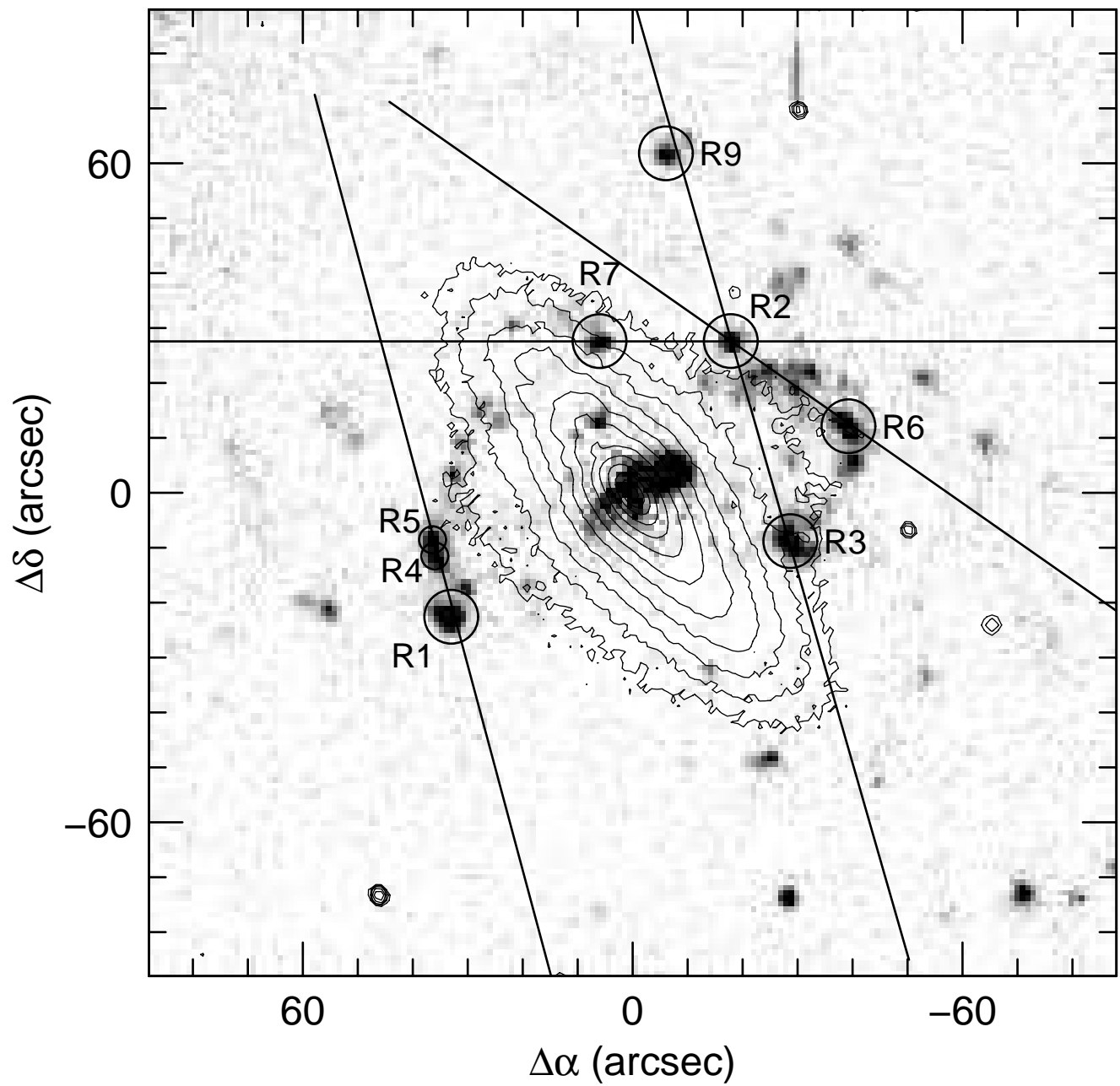
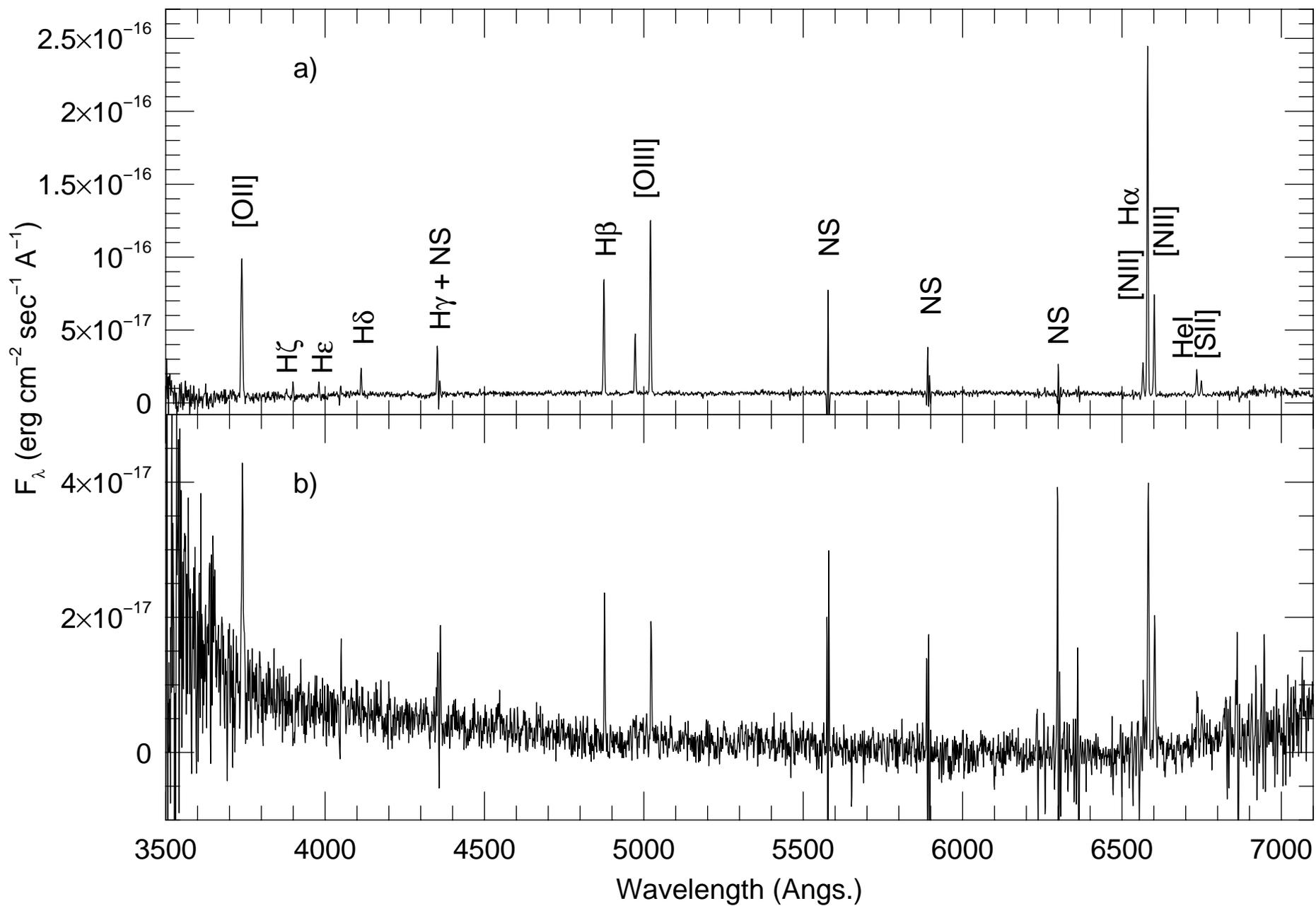
\end{document}